\documentclass[useAMS,usenatbib]{mn2e}

\usepackage{epsfig}
\usepackage{longtable}
\usepackage{times}

\usepackage{amsmath}

\def \inte {$INTEGRAL$}
\def \xmm {$XMM$-$Newton$}

\def \sw {$Swift$}

\def \src {IGR\,J16418--4532}

\def \hcm {\hbox {\ifmmode $ atom cm$^{-2}\else atom cm$^{-2}$\fi}}
\def \arcmin {\hbox{$^\prime$}}
\def \arcsec {\hbox{$^{\prime\prime}$}}
\def \chisq {$\chi ^{2}$}

\def \apj {ApJ}
\def \apjl {ApJL}
\def \apjs {ApJS}
\def \aap {A\&A}

\def \mnras {MNRAS}
\def \araa {ARA\&A}

\def \physrep {Phys. Reports}
\newcommand{\be}{\begin{equation}}

\newcommand{\ee}{\end{equation}}

\newcommand{\msun}{{~M_{\odot}}}

\title[\xmm\ observation of \src]{The \xmm\ view  of
supergiant fast X--ray transients: the case of \src}
\author[L. Sidoli, et al.]{L.\ Sidoli,$^{1}$\thanks{E-mail: sidoli@iasf-milano.inaf.it} S.\ Mereghetti,$^{1}$  V.\ Sguera,$^{2}$ and F.\ Pizzolato,$^{1}$ \\
$^{1}$INAF, Istituto di Astrofisica Spaziale e Fisica Cosmica,
	Via E.\ Bassini 15,   I-20133 Milano,  Italy \\
$^{2}$INAF, Istituto di Astrofisica Spaziale e Fisica Cosmica,
       Via Gobetti 101, I-40129 Bologna, Italy
}

\begin{document}

\date{Accepted 2011 October 23. Received 2011 October 18; in original form 2011
August 29}

\pagerange{\pageref{firstpage}--\pageref{lastpage}} \pubyear{2011}

\maketitle

\label{firstpage}

\begin{abstract}

We report  on a 40~ks long,  uninterrupted X--ray observation
of the candidate supergiant fast X--ray transient (SFXTs) \src\
performed with \xmm\  on February 23, 2011.
This high mass X--ray binary lies in the direction of the Norma
arm, at an estimated distance of 13~kpc.
During the observation, the source showed strong variability
exceeding two orders of magnitudes, never observed before from
this source. 
Its X--ray flux varied in the range from $\sim$0.1~counts~s$^{-1}$ to $\sim$15~counts~s$^{-1}$,
with several
bright flares of different durations (from a few hundreds to a
few thousands seconds) and sometimes with a quasi-periodic
behaviour. This finding supports the previous suggestion that
\src\ is a member of the SFXTs class. In our new observation we
measured a pulse period of 1212$\pm$6 s, thus confirming that
this binary contains a slowly rotating neutron star.
During the periods of  low luminosity the source spectrum is
softer and more absorbed than during the flares. A soft excess
is present below 2 keV in the cumulative flares spectrum,
possibly due to  ionized wind material at a distance similar to
the neutron star accretion radius.
The kind of X--ray variability displayed by \src, its dynamic range and time scale,
together with the sporadic presence of quasi-periodic flaring, all
are suggestive of a transitional accretion regime
between pure wind accretion and full Roche lobe overflow.
We discuss here for the first time this hypothesis to explain the behaviour of \src\ and, possibly, of other
SFXTs with short orbital periods.
\end{abstract}


	\section{Introduction\label{intro}}

\src\ is a transient X-ray source discovered in the direction
of the Norma Galactic arm during \inte\ observations performed
on 2003 February 1--5 \citep{Tomsick2004}.
Its proposed optical/infrared  counterpart (2MASS~J16415078--4532253, \citealt{Chaty2008}, \citealt{Rahoui2008}),
positionally consistent with the accurate localization obtained with the \sw\ X-Ray Telescope (\citealt{Romano2011:igr16418}),
is indicative of a high mass X--ray binary (HMXB)  at a distance of about 13~kpc.
The short time interval of X--ray activity observed with \inte\
suggested to classify \src\ as a supergiant fast X--ray transient (SFXT, \citealt{Sguera2006}).

\citet{Walter2006} reported the presence of pulsations at
1246$\pm{100}$~s in an \xmm\ observation obtained in 2004,
while  a periodicity  of 3.753$\pm{0.004}$ days
\citep{Corbet2006igr16418} or 3.7389$\pm{0.0004}$~days
(\citealt{Corbet2006igr16418}, \citealt{Levine2011}) was
obtained from an analysis of the long term light curves from
the \sw\ Burst Alert Telescope and the $RXTE$ All Sky Monitor
data (ASM), respectively. This  modulation has been interpreted
as the orbital period\footnote{the formal discrepancy between
the two periods is probably due to an underestimation of the
respecive errors.} of the binary system.

Following the MAXI satellite discovery of renewed X-ray
activity from the region of \src\ in February 2011 (as reported
in \citealt{Romano2011:igr16418}), we performed a Target of
Opportunity \xmm\ observation.
The new \xmm\ data show  variability spanning an unprecedented  range in this source,
thus  supporting its SFXT nature.
Furthermore, the frequent presence of strong quasi-periodic flares suggests that this long spin-period pulsar in a
relatively compact HMXB  (as implied by the short orbital period), might be in an intermediate accretion regime
between Roche-lobe overflow and wind accretion.

 	 \section{Observations and Data Reduction\label{dataredu}}


\src\ was observed with \xmm\  between 2011 February 23 (at
13:55 UT) and February 24 (at 00:32 UT), with a net exposure of
about 39~ks.
The \xmm\ $Observatory$ carries three 1500~cm$^2$ X--ray
telescopes, each with an European Photon Imaging Camera (EPIC)
at the focus. Two of the EPIC use Metal Oxide Semi-conductor
(MOS) CCDs \citep{Turner2001} and one uses a pn CCD
\citep{Struder2001}. Reflection Grating Spectrometer (RGS)
arrays \citep{DenHerder2001} are located behind two of the
telescopes.

Data were reprocessed using version 11.0 of the Science Analysis
Software (SAS).
Both MOS and pn operated in Full Frame mode and used the medium thickness filter.
Extraction radii of 40\arcsec\ and 1\arcmin\ were  used for the
source events, respectively for the pn and MOS cameras.
Background counts were obtained from similar sized regions,
offset from the source position, and in the same temporal
intervals, when dealing with time selected analysis.
The background (selected with PATTERN=0 and in the energy range 10--12 keV in the pn)
showed evidence of flaring activity only during the first 12~ks of the observation.
Therefore, we removed the corresponding time interval in most of our analysis, as discussed in detail in
the next section.

Response and ancillary matrix files
were generated using the SAS tasks {\sc rmfgen} and {\sc arfgen}.
Using the SAS task {\sc epatplot}, we found that MOS spectra
were affected by pile-up. Thus, we report here only on EPIC pn
spectroscopy (adding MOS data did not improve the spectral
fitting), while for the timing analysis both the two MOS and
the pn were considered.
Spectra were selected
using patterns from 0 to 4 with the pn.

To ensure applicability of the \chisq\ statistics, the
net spectra were rebinned such that at least 30 counts per
bin were present and such that the energy resolution was not
over-sampled by more than a factor 3.  All spectral
uncertainties and upper-limits are given at 90\% confidence for
one interesting parameter.
In the spectral fitting we used the photoelectric absorption
model {\sc phabs} in {\sc xspec} with the interstellar
abundances of \citet{Wilms2000}.
The RGSs were operated in spectroscopy mode \citep{DenHerder2001}, but
given the high absorbing column density, they did not detect the source.

To better investigate the long term behaviour of \src\ we
reanalysed also the 2004 \xmm/EPIC observation, first reported
by \citet{Walter2006}, with the same procedures and selection
criteria used for the 2011 data\footnote{except for the pn
source extraction region, for which we adopted a radius of
20\arcsec\ to avoid stray light contamination from a bright
transient source located outside the field of view.}.

  	\section{Analysis and Results\label{result}}

\subsection{Spectroscopy}
\label{sec:spec}

\begin{figure*}
\begin{center}
\centerline{\includegraphics[width=6.5cm,angle=-90]{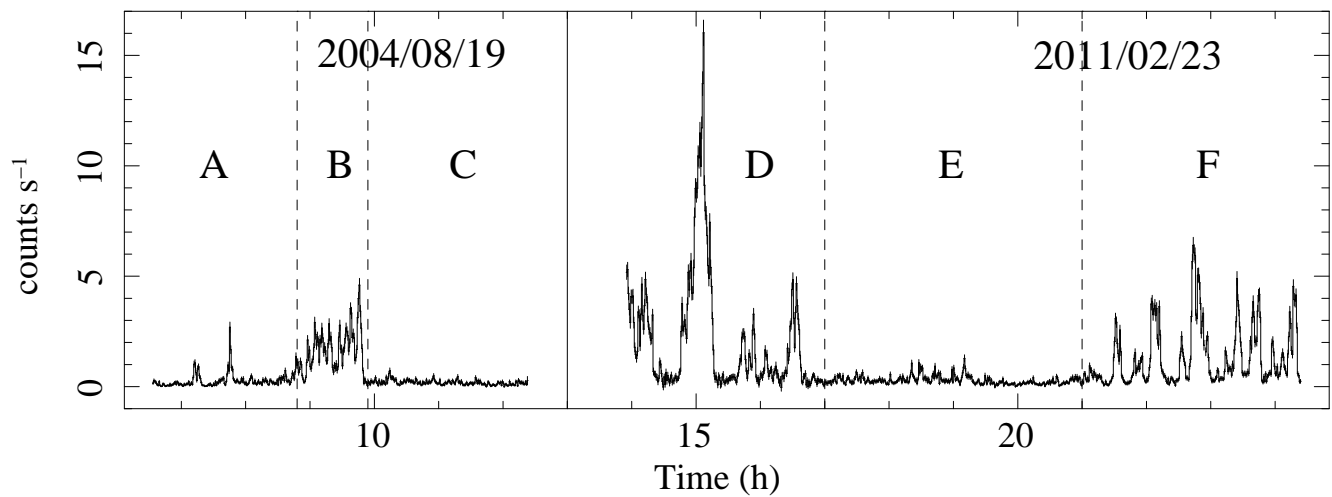}}
\caption{EPIC pn, background-subtracted, light curves of \src\
in the 0.3--12 keV energy range. The bin size is 50~s and the
time axis is in UTC hours of the dates indicated in the panels.
Letters indicate the different intervals used for the time
selected spectroscopy. Note that also during the non-flaring
time intervals, the source is significantly detected, with
average count rates 0.185$\pm{0.006}$~counts~s$^{-1}$  and
0.331$\pm{0.005}$~counts~s$^{-1}$ in 2004 (interval C) and 2011
(interval E), respectively (0.3--12 keV). } \label{fig:lc}
\end{center}
\end{figure*}

The EPIC pn, background subtracted, light curves of \src\
observed in 2004 and in 2011 are shown in Fig.~\ref{fig:lc}.
Letters mark time intervals showing  different kinds of X-ray
activity within each observation. In particular, strong flares
are seen in time intervals B, D, and F, while in the remaining
time intervals the source was in a low intensity state, with
less frequent and fainter flares. As mentioned above, the first
part of the 2011 observation (time interval D) was affected by
high background level, so in the following we will not use this
time interval, except when extracting the spectra from the
peaks of the bright flares (for which the background
contribution is negligible).

\begin{figure}
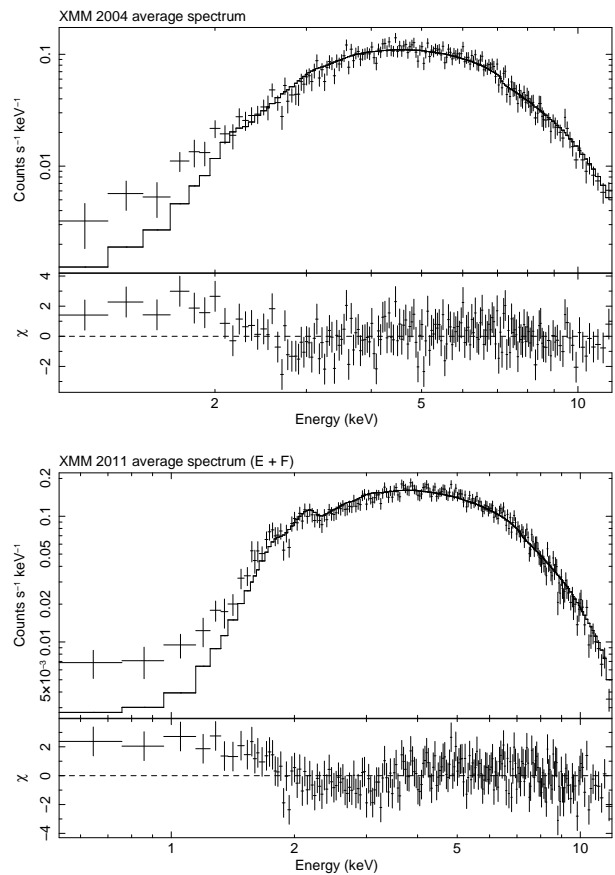

\begin{center}
\begin{tabular}{cc}
\includegraphics[height=8.0cm,angle=-90]{ldadelchi_pow_2004_average.ps} \\
\includegraphics[height=8.0cm,angle=-90]{ldadelchi_average_2011_spec_pow.ps}
\end{tabular}
\end{center}
\caption{Spectral results of the average spectra extracted from the EPIC pn observations
in 2004 ({\em upper panel}) and in 2011 ({\em lower panel}). Count spectra are shown, together
with the residuals (in units of standard deviations) of the data to
the absorbed power law model (Table~\ref{tab:av_spec}).
}
\label{fig:av_spec}
\end{figure}

We first analysed the average spectrum of  each observation,
corresponding to a net integration time of 19.6~ks for the 2004
data, and  23.8~ks for the 2011 data (intervals E and F). The
best fit parameters obtained with an absorbed power law model
are listed in Table~\ref{tab:av_spec} and  the spectra shown in
Fig.~\ref{fig:av_spec}. A similar spectral slope is seen in the
two observations,  while the absorption in 2011 was about a
half of that observed in 2004.
While the  2004 spectrum is reasonably well fitted by the power
law model (although with marginal evidence for a soft excess
below 2 keV),  structured wave-like residuals over the entire
energy range are evident in the 2011 spectrum, with a positive
excess below 2 keV more pronounced than in 2004.

\begin{table}
\begin{center}
\caption[]{Spectral results of the time averaged spectra of the
2004 and 2011 \xmm\ observations. An absorbed power law model
was used. $\Gamma$ is the power law photon index. Flux is in
the 1--10~keV energy range in units of
10$^{-11}$~erg~cm$^{-2}$~s$^{-1}$ and is corrected for the
absorption, N$_{\rm H}$ (in units of $10^{22}$~cm$^{-2}$). }
\begin{tabular}{llll}
 \hline
\hline
\noalign {\smallskip}
Parameter   &           2004    (A+B+C)            &     2011 (E+F)                    \\
\hline
\noalign {\smallskip}
N$_{\rm H}$             &  $ 15 \pm{1}$             &     $7.4 \pm{0.3}$              \\
$\Gamma$                &  $1.3 \pm{0.1}$           &     $1.28 \pm{0.05}$              \\
Unabs. Flux             &    1.8                    &     1.5                          \\
$\chi^{2}_{\nu}$/dof    &    1.147/166              &     1.252/210                    \\
\noalign {\smallskip}
\hline
\label{tab:av_spec}
\end{tabular}
\end{center}
\end{table}

We next considered temporal selected spectra, 
extracting different spectra from the flaring intervals and from the low intensity emission, 
within each observation.


We first considered the 2004 observation, when  \src\ showed a
bright flare with a complex shape (interval B,
Fig.~\ref{fig:lc}, left panel) in the middle of two
low-intensity time intervals (A and C).
We extracted a spectrum from each time interval (A, B and C),
with the one of interval B formed by selecting only the events
corresponding to time intervals with count rate above
0.5~counts~s$^{-1}$.
The results for these spectra, fitted with an absorbed power
law, are reported in Table~\ref{tab:spec}. We also extracted a
further low intensity spectrum from time intervals  A and C,
but now excluding the two faint flares that occurred in time
interval A. This ``cleaned A+C'' spectrum is indeed more
representative of the low intensity emission in 2004. A
comparison with the flare spectrum (B) indicates that the \src\
X--ray emission is less absorbed during flares. There is also a
possible hint for a flatter spectral slope when the source is
brighter, but the uncertainties are too large to draw a firm
conclusion, based on the 2004 observation only.

\begin{table*}
\begin{center}
\caption[]{Results of the  time selected spectroscopy (letters
mark the same time intervals displayed in Fig.~1). An absorbed
power law model was used. Flux is in the 1--10~keV energy range
in units of 10$^{-11}$~erg~cm$^{-2}$~s$^{-1}$ and is corrected
for the  absorption, N$_{\rm H}$ (in units of
$10^{22}$~cm$^{-2}$). }
\begin{tabular}{lllllll}
 \hline
\hline
\noalign {\smallskip}
Parameter   &           A                           &     B                         &    C                           &       A + C                    &        E                    &   F    \\
            &                                       &                               &                                &      (cleaned)               &                             &          \\  
\hline
\noalign {\smallskip}
N$_{\rm H}$             &  $ 20 ^{+3} _{-3}$        &    $13  ^{+1} _{-1}$          &    $17  ^{+4} _{-2}$           &     $ 19 ^{+2} _{-2}$         &    $8.2  ^{+0.8} _{-0.6}$   &   $6.72  ^{+0.36} _{-0.34}$     \\
$\Gamma$                &  $1.25 ^{+0.25} _{-0.23}$ &    $1.29  ^{+0.11} _{-0.11}$  &    $1.47  ^{+0.35} _{-0.23}$   &     $1.54 ^{+0.17} _{-0.22}$  &    $1.57 ^{+0.13} _{-0.10}$ &   $1.11  ^{+0.06} _{-0.06}$     \\
Unabs. Flux             &    1.1                    &     6.5                       &    0.7                         &      0.8                      &     0.65                    &    4.4  \\
$\chi^{2}_{\nu}$/dof    &    1.035/41               &     1.130/128                 &    0.690/35                    &      1.085/60                 &     1.008/90                &    1.196/176   \\
\noalign {\smallskip}
\hline
\label{tab:spec}
\end{tabular}
\end{center}
\end{table*}

We next considered the observation performed in 2011.
In order to investigate the spectral properties of the single
flares we selected all time intervals with a count rate above
1~counts~s$^{-1}$ in the pn light curve (0.3--12 keV) binned at
256~s. This time selection yielded 16 single flares (6 flares
in time interval D and 10 in  F).
%
\begin{figure}
\begin{center}
\begin{tabular}{cccc}
\includegraphics[height=5.cm,width=8.5cm,angle=0]{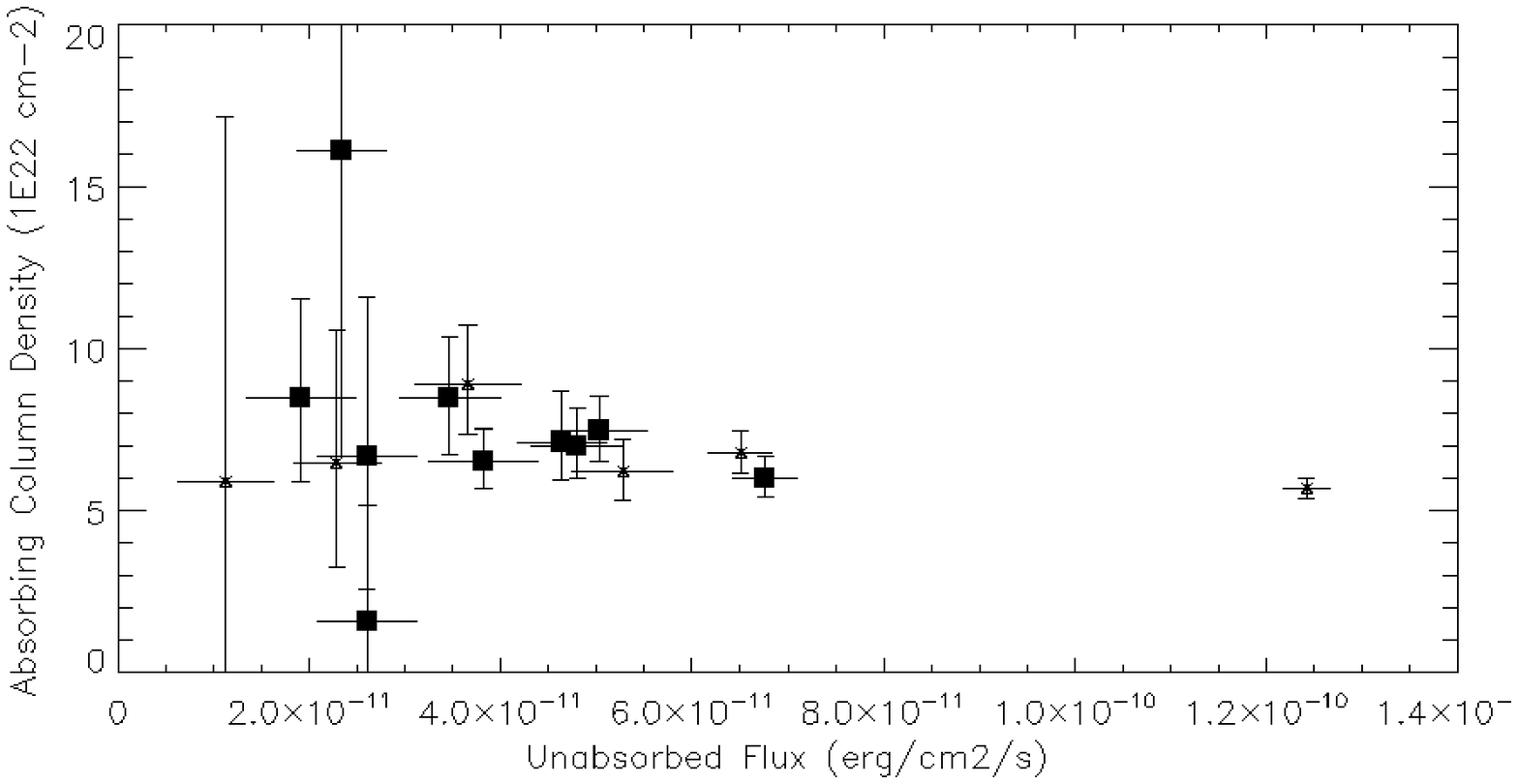} \\
\includegraphics[height=5.cm,width=8.5cm,angle=0]{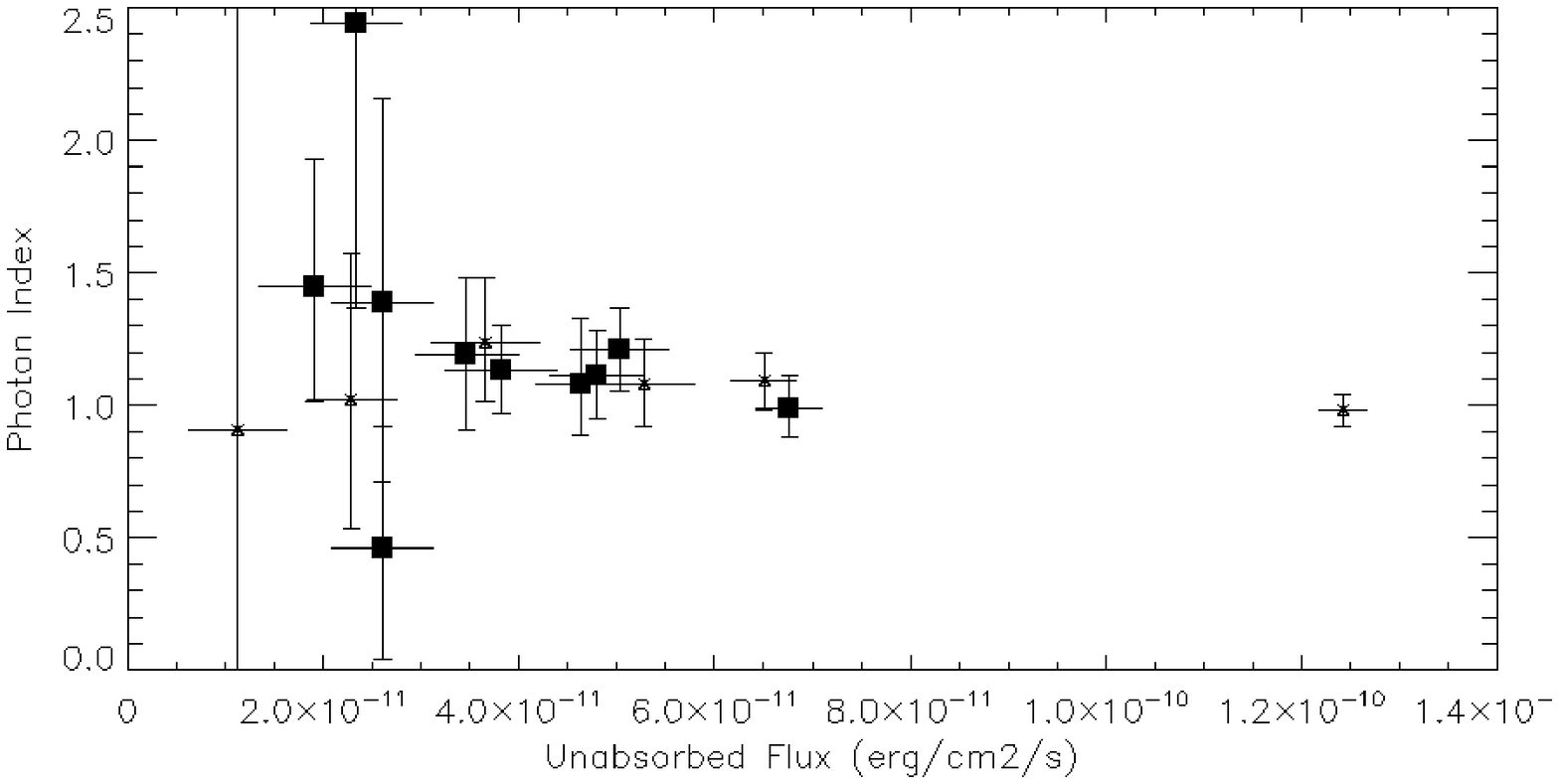}
\end{tabular}
\end{center}
\caption{Spectral results of the intensity selected spectroscopy of all flares observed in 2011.
The assumed model is an absorbed power law.
{\em Upper panel} shows the absorbing column density evolution with the flux corrected for the
absorption, while the {\em Lower panel} displays the photon index versus unabsorbed flux (1--10 keV).
{\em Light triangles} mark the spectral parameters of the flares extracted from the
first part of the 2011 observation (interval D in Fig.~1), while {\em solid black squares}
indicate flares occurred in the second flaring part of the same observation (interval F in Fig.~1).}
\label{fig:flarepowspec}
\end{figure}
%
%
An absorbed power-law gave
almost always a good fit to each single flare spectrum. 
The best fit parameters of each flare  are shown in
Fig.~\ref{fig:flarepowspec}, which shows  no evidence for
significant spectral variability between the flares, even
comparing flares in the first part of the observation (interval
D) with those located in the second part (interval F). Most of
the flare spectra require an absorption  well in excess of the
total Galactic value towards \src\
(1.9$\times$10$^{22}$~cm$^{-2}$, \citealt{DL90}). A black body
model always results in a worse (or in an equally good) fit,
giving  densities, on average, a half of the value obtained
with a power law. The black body temperatures are $\sim$2~keV,
and the resulting black body radii have values of a few hundred
meters at 13~kpc.

\begin{figure}
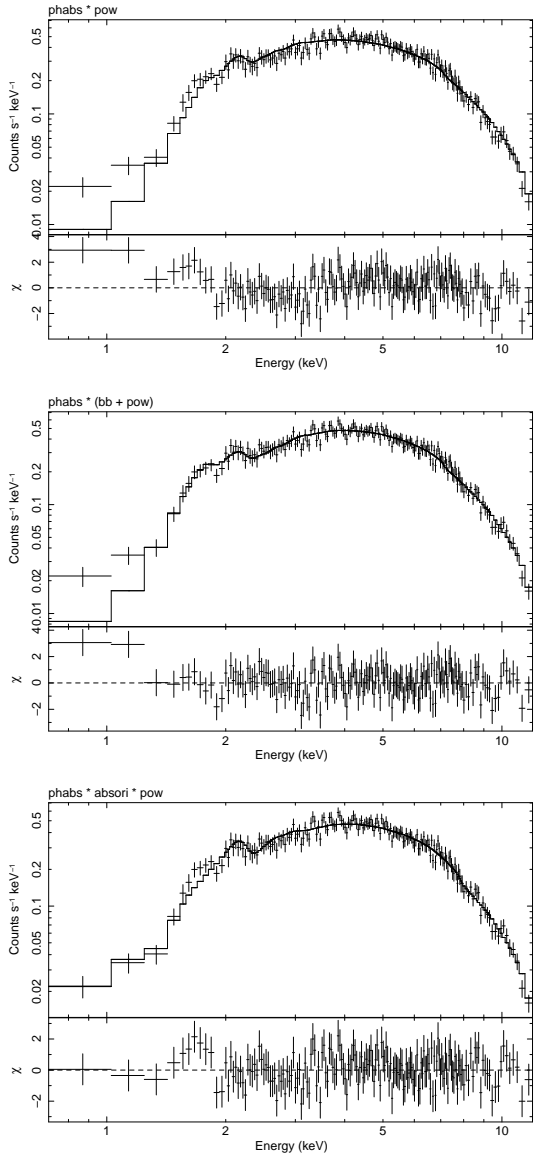

\begin{center}
\begin{tabular}{cccc}
\includegraphics[height=7.0cm,angle=-90]{ldadelchi_allflares_secondpart_pow.ps} \\
\includegraphics[height=7.0cm,angle=-90]{ldadelchi_flares_secondpart_bb_pow.ps} \\
\includegraphics[height=7.0cm,angle=-90]{ldadelchi_allflares_secondpart_absori_pow.ps}
\end{tabular}
\end{center}
\caption{Cumulative X--ray spectrum of the flaring activity of the second part of the 2011  observation
(flares in interval F).
The {\em upper panel} shows the counts spectrum together with the residuals (in units of standard deviations)
when fitting with an absorbed power law.
The {\em middle panel} shows the residuals when fitting with a power law together with a black body.
{\em Lower panel} displays the residuals when fitting with a power law continuum modified by a
ionized absorber ({\sc absori} in {\sc xspec}, see Sect.\ref{sec:spec} for details).
}
\label{fig:allflares_second}
\end{figure}

We next extracted a cumulative ``spectrum during flares'' by
summing all  the spectra of the flares occurring in the time
interval  F, reaching a net exposure time of 5085~s.
An absorbed power law gave a good fit  above 2~keV, but showed
some positive residuals (although never exceeding 3~$\sigma$
from the model) at softer energies (reduced
$\chi^{2}_{\nu}$=1.196 for 176 degrees of freedom, dof, see the upper panel in
Fig.~\ref{fig:allflares_second}).
These positive residuals suggest the presence of either (1) an
additional soft component, (2) an ionized absorber or (3) a
partial covering.
Thus, we first added a black body component to the power law
continuum, obtaining  a better fit  (reduced $\chi^2$=0.987 for
174 dof; F-test probability of 5.53$\times$10$^{-8}$). This
resulted in a softer and more absorbed power law, and a black
body with temperature of 0.17~keV and emitting radius of a few
hundred km.

The use of an ionized absorber ({\sc absori} in {\sc xspec};
\citealt{Done1992}), in addition to a neutral one accounting
for the interstellar absorption ({\sc phabs}), also resulted in
a better fit with respect to the simple absorbed power law
(reduced $\chi^2$=0.950 for 174 dof; F-test probability of
1.99$\times$10$^{-9}$). The absorber temperature was fixed to
30~kK and the power law photon index of the photoionizing
source was linked to the photon index of the continuum. The
absorber ionization state $\xi$ ~(defined as $\xi$=L/nR$^2$,
where L is the X--ray luminosity of the ionizing source, n is
the density of the absorber, and R is the distance of the
ionized matter from the X--ray source), resulted in a value of
125~$^{+60} _{-45}$~erg~cm~s$^{-1}$.

A better fit with respect to the simple absorbed power law
(reduced $\chi^{2}_{\nu}$=0.971 for 174 dof; F-test probability
of 1.335$\times$10$^{-8}$) was obtained also using a power law
continuum modified by a partial covering  absorption ({\sc
pcfabs} in  {\sc xspec}), together with the usual absorption
accounting for the interstellar value. Partial covering
parameters are the equivalent hydrogen column density N$_{\rm H
pcfabs}$ and the dimensionless covering fraction, f (${\rm 0 <
f < 1}$).
All the best fit parameters of these models are summarized in
Table~\ref{tab:allflares_second}.
The counts spectra
together with their residuals when fitting with  three
different continua are reported in
Fig.~\ref{fig:allflares_second}. The residuals of the partial
covering model are not shown because they look very similar to
those obtained with the additional soft component (middle panel
in Fig.~\ref{fig:allflares_second}).
There is no evidence  of an iron emission line. Upper limits
(95\% confidence level) to the equivalent width (EW) of a
narrow  line from ionized iron (energy fixed at 6.7~keV) can be
placed at EW$<$34~eV, while for a neutral iron emission line
(energy fixed at 6.4~keV) at EW$<$16~eV, assuming a power law
continuum.
%

\begin{table}
\begin{center}
\caption[]{Spectral results for the cumulative spectrum during
flares in time interval F (2011) fitted with  models able to
account for the soft excess. Flux is in the 1--10~keV energy
range in units of 10$^{-11}$~erg~cm$^{-2}$~s$^{-1}$ and is
corrected only for the absorption N$_{\rm H}$ ({\sc phabs} in
{\sc xspec}). Absorbing column densities are in units of
$10^{22}$~cm$^{-2}$. }
\begin{tabular}{llll}
 \hline
\hline
Parameter                          &        {\sc  pow + bb    }          &      {\sc  absori * pow }  &        {\sc  pcfabs * pow }                       \\
\hline
\noalign {\smallskip}
N$_{\rm H}$        &    $9.41  ^{+0.48} _{-1.32}$  &    $0.65  ^{+0.70} _{-0.50}$   &      $5.39 ^{+0.69} _{-0.86}$     \\
N$_{\rm H absori}$ &                 $-$           &     $15.8  ^{+2.0} _{-1.8}$    &                      $-$          \\
$\xi$              &                 $-$           &     $125  ^{+60} _{-45}$       &                      $-$          \\
N$_{\rm H pcfabs}$ &                 $-$           &         $-$                    &      $11.7  ^{+4.8} _{-3.7}$      \\
f                  &                 $-$           &         $-$                    &      $0.59  ^{+0.12} _{-0.11}$    \\
$\Gamma$           &    $1.33  ^{+0.08} _{-0.12}$  &    $1.29  ^{+0.07} _{-0.07}$   &      $1.42  ^{+0.12} _{-0.12}$    \\
kT$_{{\rm bb}}$ (keV) &  $0.17  ^{+0.02} _{-0.03}$ &     $-$                        &         $-$                       \\
R$_{\rm bb}$    (km)  &  $250  ^{+430} _{-90}$     &     $-$                        &         $-$                       \\
Unabs. Flux           &     10.7                      &    3.1                         &      4.0                          \\
$\chi^{2}_{\nu}$/dof  &     0.987/174                 &    0.950/174                   &     0.971/174                     \\
\noalign {\smallskip}
\hline
\label{tab:allflares_second}
\end{tabular}
\end{center}
\end{table}

\begin{figure}
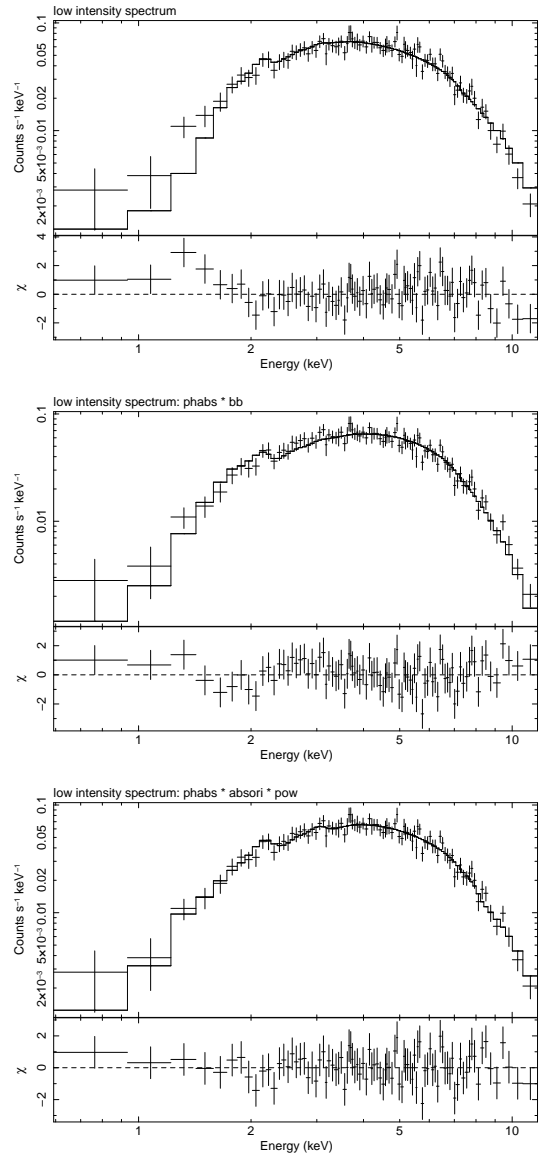

\begin{center}
\begin{tabular}{cccc}
\includegraphics[height=7.0cm,angle=-90]{ldadelchi_low_pow.ps} \\
\includegraphics[height=7.0cm,angle=-90]{ldadelchi_low_bb.ps} \\
\includegraphics[height=7.0cm,angle=-90]{ldadelchi_low_absori_pow.ps}
\end{tabular}
\end{center}
\caption{X--ray counts spectra  and residuals (in units of standard
deviations)  for  the low intensity emission observed in 2011 (interval E) fitted with an
absorbed power law model ({\em upper panel}),
an absorbed black body ({\em middle panel}), and
an absorbed power law modified with a ionized absorber ({\em lower panel}).
}
\label{fig:lowspec}
\end{figure}

The overall flares spectrum can be compared with the cumulative
spectrum extracted from the 2011 low intensity state (time
interval E, average count rate of
0.330$\pm{0.005}$~counts~s$^{-1}$), corresponding to a net EPIC
pn exposure of 14.4~ks. An absorbed power law gave a good fit
($\chi^2_{\nu}=1.008$ for 90 dof) with the following
parameters: $N_{\rm H}$=(8.2$ ^{+0.8}
_{-0.6})$$\times$$10^{22}$~cm$^{-2}$, $\Gamma$=1.57$ ^{+0.13}
_{-0.10} $. The average observed flux was
3.7$\times$$10^{-12}$~erg~cm$^{-2}$~s$^{-1}$, while the flux
corrected for the absorption was
6.5$\times$$10^{-12}$~erg~cm$^{-2}$~s$^{-1}$ (1--10~keV),
translating into an average luminosity of
1.2$\times$$10^{35}$~erg~s$^{-1}$ at 13~kpc.
Also an absorbed black body gave a good fit
($\chi^2_{\nu}=0.924$ for 90 dof), with  $N_{\rm H}$, of 3.6$
\pm{0.4}$$\times$$10^{22}$~cm$^{-2}$, a temperature, kT$_{\rm
bb}$, of 1.78$\pm{0.07}$~keV and a radius, R$_{\rm bb}$, of
0.29$\pm{0.02}$~km (at 13~kpc).
Finally, we fitted the low intensity spectrum with the best fit
model obtained for the flares, i.e. an absorbed power law
modified with a ionized absorber. This improved the  fit
quality ($\chi^2_{\nu}=0.794$ for 88 dof, F-test probability of
2.75$\times$10$^{-5}$) and gave the following parameters:
$N_{\rm H}$=(4$\pm{2}$)$\times$$10^{22}$~cm$^{-2}$,
$\Gamma$=1.85 $ ^{+0.20} _{-0.13} $, ionized absorption,
$N_{\rm H absori}$, of (2.5 $^{+0.8} _{-0.6}$)
$\times$$10^{23}$~cm$^{-2}$, ionization parameter, $\xi$, of
600  $^{+400} _{-335}$~erg~cm~s$^{-1}$.
If we fix the ionized absorption parameters
 $N_{\rm H absori}$ and  $\xi$
values obtained for the cumulative flares spectrum, we again
obtain a good fit to the low intensity spectrum
($\chi^2_{\nu}=0.886$ for 88 dof), resulting in a column
density $N_{\rm H}$=(2.7 $^{+0.7}
_{-0.6}$)$\times$$10^{22}$~cm$^{-2}$ and in a power law photon
index $\Gamma$=1.7$\pm{0.1}$, consistent with previous results.
Therefore, we   conclude that the ionized absorber found in the
overall flaring spectrum is consistent with  being  present
also during the low intensity emission with similar properties,
although the statistics does not permit to be very sensitive to
its parameters. In any case, the low intensity emission is
softer than the flaring spectrum.
The low intensity spectrum fitted with these models is shown in
Fig.~\ref{fig:lowspec}.

Finally, we also performed an intensity selected spectroscopy: 
a series of intensity selected spectra were produced, using intervals corresponding to EPIC pn count rates
of $<$1, 1--5, 5--10, and $>$10 
counts~s$^{-1}$, when the data
are accumulated with a binning of 256~s, for both observations (2004 and 2011).
The best-fit, obtained with an absorbed power law, is  reported in Table~\ref{tab:intsel},
and it clearly indicates that the spectrum is harder when the source is brighter, as usually
observed in accreting pulsars.

\begin{table}
\begin{center}
\caption[]{Results of the  intensity selected spectroscopy.
An absorbed
power law model was used. Flux is in the 1--10~keV energy range
in units of 10$^{-11}$~erg~cm$^{-2}$~s$^{-1}$ and is corrected
for the  absorption, N$_{\rm H}$ (in units of
$10^{22}$~cm$^{-2}$). Uncertainties on the unabsorbed fluxes are about 3\%.}
\begin{tabular}{lllll}
 \hline
\hline
\noalign {\smallskip}
Obs. 2004             &      $<$1   c~s$^{-1}$     &     1--5       c~s$^{-1}$   &    5--10  c~s$^{-1}$    &      $>$10  c~s$^{-1}$             \\
Parameter             &                                  &                                   &                                &                                             \\  
\hline
\noalign {\smallskip}
N$_{\rm H}$             &  $ 19 ^{+2} _{-2}$        &    $13  ^{+1} _{-1}$          &   &            \\
$\Gamma$                &  $1.54 ^{+0.17} _{-0.22}$ &    $1.29  ^{+0.11} _{-0.11}$  &   &            \\
Unabs. Flux             &    0.8                    &     6.5                       &   &            \\
$\chi^{2}_{\nu}$/dof    &    1.085/60               &     1.130/128                 &   &            \\
\hline
\noalign {\smallskip}
Obs. 2011            &     $<$1   c~s$^{-1}$     &     1--5       c~s$^{-1}$   &    5--10  c~s$^{-1}$    &      $>$10  c~s$^{-1}$             \\
Parameter            &                           &                             &                         &                         \\  
\hline
\noalign {\smallskip}
N$_{\rm H}$             &  $ 8.2 ^{+0.8} _{-0.6}$        &  $6.9  ^{+0.3} _{-0.3}$  &    $5.70 ^{+0.53} _{-0.49}$     &     $ 5.60 ^{+0.50} _{-0.50}$          \\
$\Gamma$                &  $1.57 ^{+0.13} _{-0.10}$ &    $1.12  ^{+0.05} _{-0.05}$  &    $1.05  ^{+0.10} _{-0.10}$    &     $0.88 ^{+0.09} _{-0.09}$   \\
Unabs. Flux             &    0.7                    &     5.0                       &    11.7                         &      20.0                       \\
$\chi^{2}_{\nu}$/dof    &    1.008/90               &     1.147/209                 &    0.823/119                    &      1.190/137                  \\
\noalign {\smallskip}
\hline
\label{tab:intsel}
\end{tabular}
\end{center}
\end{table}

\subsection{Timing Analysis}
\label{sec:timing}

\begin{figure}
\begin{center}
\centerline{\includegraphics[width=6.8cm,angle=-90]{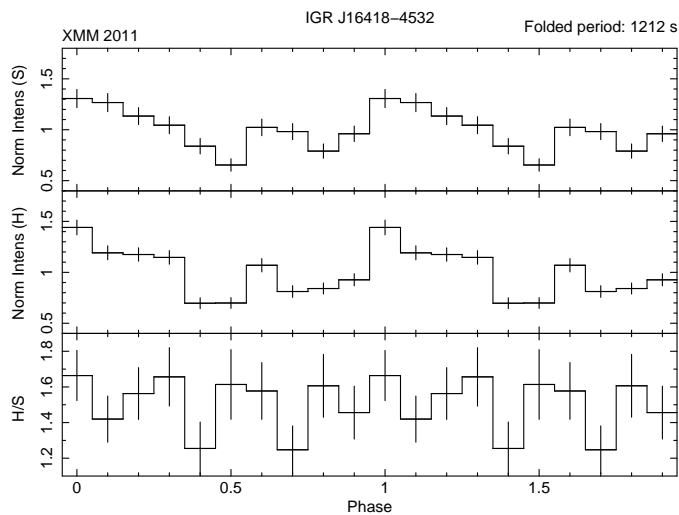}}
\caption{\src\ pulse profiles at soft (0.3--4 keV) and hard energies (4--12 keV)
with \xmm\ in 2011 (only the low intensity emission), together with their hardness ratio, obtained
folding the light curve at a period of 1212~s. The zero phase is arbitrary.}
\label{folded2011}
\end{center}
\end{figure}

For the timing analysis we  used both MOS and pn
EPIC data. Arrival times were corrected to the Solar System
barycenter, but not for the 3.7 days orbital motion, owing to the unknown parameters of the system.
The observability of the periodicity at 1246$\pm{100}$~s \citep{Walter2006} is complicated by the
presence of the strong flares which occur on a comparable time scale.
Folding the data of the whole observation at different trial periods produces several
peaks in the $\chi^2$ distribution.
On the other hand, by restricting the period search to the
central part of the 2011 observation (interval E), when the
source was in a low intensity and non-flaring state, the X--ray
pulsations are clearly detected at a period, P, of
1212$\pm{6}$~s ($\chi^2$=252, for 9 dof). The corresponding folded light curve is plotted
in Fig.~\ref{folded2011}.

\begin{figure}
\begin{center}
\centerline{\includegraphics[width=6.8cm,angle=-90]{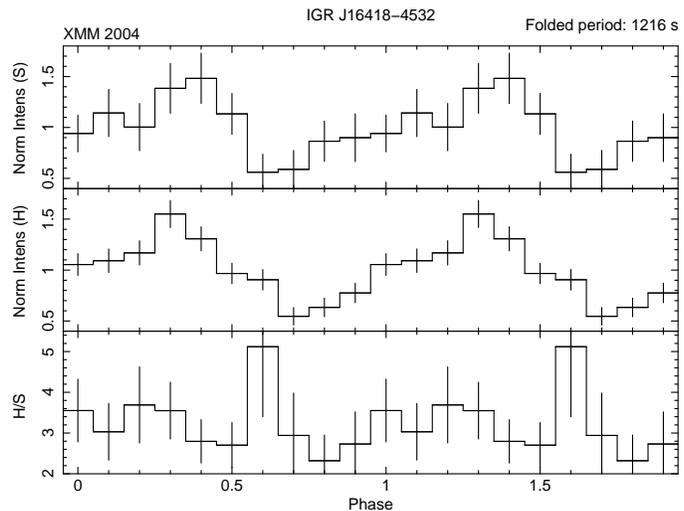}}
\caption{\src\ pulse profiles at soft (0.3--4 keV) and hard energies (4--12 keV)
with \xmm\ in 2004 (only the low intensity emission), together with their hardness ratio, obtained
folding the light curve at a period of 1216~s. The zero phase is arbitrary.}
\label{folded2004}
\end{center}
\end{figure}

Also in the 2004 observation the source pulsations are more
easily detected during the non flaring time interval. By a
standard folding analysis of the time interval C
we obtained a clear peak in the $\chi^2$ distribution
at  P = 1216$\pm$7~s  ($\chi^2$=257 for 9 dof). 
Based on  this result, we could then
refine the period estimate using the data from the whole
observation, which gave P = 1213$\pm{2}$~s. The folded light
curve for the non-flaring time interval is plotted in
Fig.~\ref{folded2004}. Contrary to that obtained in 2011, it
shows a single broad peak.

We finally performed a timing analysis of a series of $RXTE$/PCA archival
observations carried out between 2009 April 5 and April 15,
obtaining a best fit spin period of 1209.4$\pm{1}$~s.
The comparison of the pulse periods measured with \xmm\ in 2004
and 2011 and with $RXTE$ in 2009 does not show significant
spin-up or spin-down.

	\section{Discussion \label{discussion}}

\inte\ showed that \src\ is an X--ray transient with frequent
flaring activity: it detected source outbursts above
10$^{-10}$~erg~cm$^{-2}$~s$^{-1}$ (20--100 keV) for about 1\%
of the \inte\ total exposure time of the source region
\citep{Ducci2010}.
This led to suggest \src\ as a member of the class of the
supergiant fast X--ray transients (SFXTs, \citet{Sguera2006};
see \citet{Sidoli2011texas} for an updated review of the SFXTs
properties).
It has been classified as a ``candidate'' SFXT because the
supergiant nature of its optical counterpart has not been
confirmed yet, and the X--ray dynamic range was not as extreme
as the prototypical SFXTs (typically, from 3 to 5 orders of
magnitudes in X--ray intensity). Among SFXTs, \src\ is, after
IGR~J16479--4514 and IGR~J18483--0311, the source with the most
frequent bright flaring activity (\citealt{Ducci2010},
\citealt{Sidoli2011texas}).

Our \xmm\ observation allowed us to continuously follow the
source variability with unprecedented detail  and to reveal,
for the first time in this source, a wide dynamic range,
typical of other so-called ``intermediate'' SFXTs, like e.g.
IGR~J18483--0311 \citep{Sguera2007} or IGR~J16465-4507
\citep{Clark2010}.

\begin{figure}
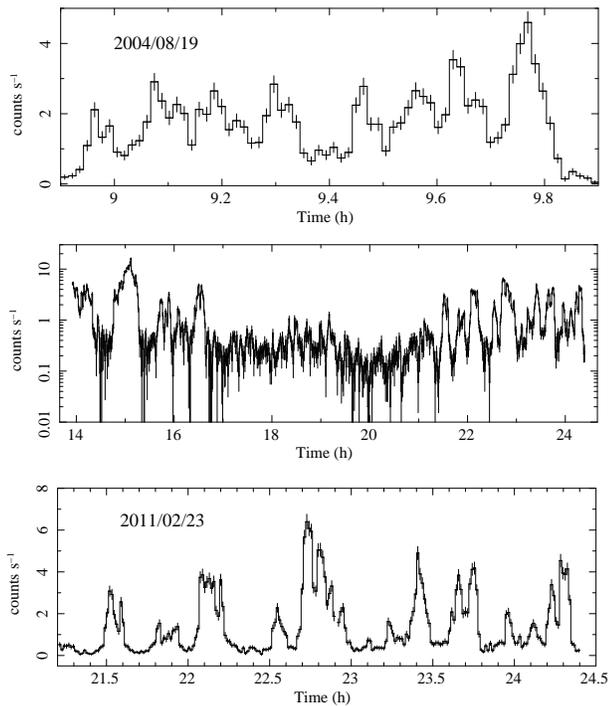

\begin{center}
\begin{tabular}{ccc}
\includegraphics[width=2.9cm,angle=-90]{lc_zoom_qdp_2004.ps} \\
\hspace{-0.1cm}
\includegraphics[width=2.9cm,angle=-90]{lc2011_log.ps} \\
\hspace{0.2cm}
\includegraphics[width=3.0cm,angle=-90]{lc_zoom_qdp_2011.ps}
\end{tabular}
\end{center}
\caption{Close-up views of \src\ light curves shown in Fig.~1.
In the {\em middle panel} we show the 2011 light curve (same as in Fig.~1) in logarithmic
scale to more clearly  show the high dynamic range now observed in \src.
In the  {\em upper} and {\em lower panels} two close-up views of the
2004 and 2011 observations are displayed, respectively,
to better show the time intervals where
quasi-periodic flaring activity is present.}
\label{fig:lczoom}
\end{figure}

The X--ray light curve observed in 2011 showed two episodes of
bright flaring activity (exceeding 10~counts~s$^{-1}$  in the first part  of the exposure),
separated by an interval of low intensity emission, where the
source faded to less than $\sim$0.1~counts~s$^{-1}$
(Fig.~\ref{fig:lczoom}, middle panel). The two  flaring
intervals (D and F) were characterized by different variability
patterns: in the first one the flux reached the peak during the
longest flare ($\sim$4000~s) and  showed  a variety of flare
shapes, while in interval F the flares had a more regular
appearance, with a hint of a  quasi--periodic pattern of a few
ks, not related with the long spin period
(see the lower panel of Fig.~\ref{fig:lczoom}).
A quasi--periodic flaring was also present in the central part
of the 2004  observation (time interval B in Fig.~1). As
visible in the upper panel of Fig.~\ref{fig:lczoom}, the
variability consisted of two sub-sets of four flares each,
recurring every $\sim$380--400~s.
%

Quasi--periodic flares have   been seen also in other HMXBs,
including some  SFXTs (e.g. XTE~J1739--302,
see \citealt{Ducci2010}). 
For example, in the Be/X-ray transient
EXO~2030+375 flares recurring on timescales of
3.96$\pm{0.04}$~hr, and intensity oscillations with periods of
900-1200~s, were observed and explained  with the formation and
disruption of a temporary and transient accretion disk around
the neutron star \citep{Parmar1989}. Although in EXO~2030+375
both the flare shapes (well described by a fast rise and an
exponential decay) and the time recurrence were more regular,
it is possible that a similar mechanism is responsible for the
\src\ flaring activity.
 The model proposed by \citet{Taam1988} predicts
that similar
recurrence time scales for the flares can be well reproduced by 
a relative velocity of the  neutron star with respect to the outflowing wind 
of about 1200--1400~km~s$^{-1}$ 
(see also \citealt{Ducci2010}  and references therein).

We note a striking similarity between the \src\ light curve
observed in 2011 and the prediction of hydrodynamic simulations
for systems accreting in the so-called ``transitional case''
between ``pure'' wind accretion and ``full'' Roche lobe overflow
\citep{Blondin1997}.
According to these simulations, in this particular transitional
accretion regime, when the mass donor is close to fill its
Roche lobe, the HMXB can display an X--ray light curve very
similar to what we observed in \src, both for what concerns the
variability  time scale and its dynamic range (see
Fig.~\ref{fig:lczoom}, middle panel).
The mass loss from the supergiant is dominated by the strong
wind, but with the additional contribution of a weak tidal gas
stream, focussed towards the neutron star (hereafter, NS). This
mechanism produces extreme variations in the mass accretion
rate, mainly due to the dynamical interaction of the weak tidal
gas stream with the accretion bow shock around the NS
\citep{Blondin1997}.
We are aware that these  simulations have been performed in two
dimensions (computing the gas flow within the orbital plane)
and the real 3D case is much more complex, but nevertheless we
think that the similarity with the observed light curve
deserves attention and further investigation.

Similarly, \citet{Bhattacharya1991} discussed the case of HMXBs
with  short orbital periods (P$\le$3-4.5 days), such as \src,
as sources where the orbits are typically narrow making
accretion via  ``beginning atmospheric Roche-lobe overflow'' an
attainable mechanism to produce X--ray emission.
In general, under the assumption that a supergiant star has a
sharply  defined radius (e.g. the photospheric radius), Roche
lobe overflow  starts as soon as such radius extends beyond the
Roche lobe. However, in reality the supergiant star does not
have a sharp edge or radius and above its photosphere there is
still atmospheric material in the form of strong stellar wind.
Consequently, already before the photospheric radius reaches
the Roche lobe, a small part of the stellar wind can begin to
flow towards the neutron star through the inner Lagrangian
point. In principle, this could also favour the formation of a
transient and temporary accretion disk whose disruption could
produce the observed quasi-periodic flares from \src.

\begin{figure}
\begin{center}
\centerline{\includegraphics[width=8.6cm,angle=0]{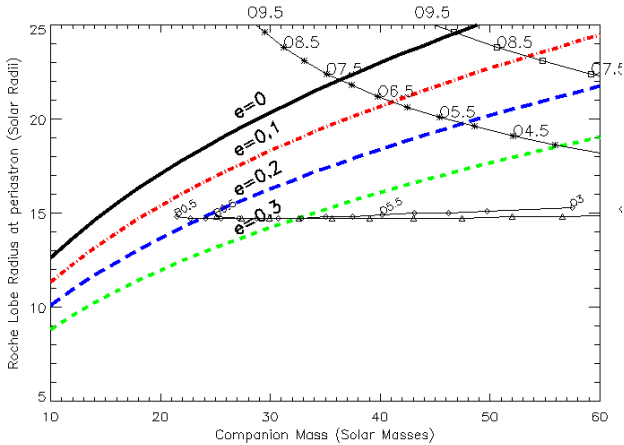}}
\caption{Roche lobe radius at periastron versus the companion mass, assuming an orbital period of 3.7389~days,
and different eccentricities.
We overplot also the  radii of early type stars, taken from \citet{Vacca1996}:
diamonds and triangles
mark OB star with luminosity class III,  their
spectroscopic and evolutionary masses, respectively
(Table~6 in Vacca et al. 1996),
while asterisks and open squares mark supergiant stars
(data are from Table~7 in Vacca et al. 1996),
their  spectroscopic and evolutionary masses, respectively,
for different spectral types. }
\label{fig:roche}
\end{center}
\end{figure}

Given the short orbital period of  3.7389~days, we plot in
Fig.~\ref{fig:roche} the Roche lobe radius of the companion
star at periastron,  as a function of its mass and for
different system eccentricities \citep{Eggleton1983}. The radii
of massive early type stars of different luminosity classes
\citep{Vacca1996} are also plotted for comparison.
The 32.8~kK best fit effective temperature\footnote{with very
large uncertainties: T$_{\rm eff}$ can range from 36 kK to 10.6
kK} of the candidate counterpart \citep{Rahoui2008}
indicates a O8.5--O9.5  type star, which can be in a narrow
circular orbit with the NS, avoiding Roche lobe overflow.  For
example, a O9~Ia (O8.5~Ia) star with a mass of 46~$\msun$
(50~$\msun$) and a radius of 24~R$_{\odot}$ \citep{Vacca1996}
in a circular orbit is consistent with the optical and X--ray
observations. Another viable possibility is  a O9.5~III
companion with a mass of 24~$\msun$ and a radius of
$\sim$15~R$_{\odot}$. In this case the eccentricity can range
from e=0 to e$\sim$0.2 without exceeding the Roche lobe
surface.
We are aware that the above speculations on the  optical
counterpart are rather uncertain in the lack of an optical
spectrum, but we here only note that an early type supergiant
(and a SFXT nature) is compatible with the short orbital
period.
Moreover, a narrow orbit in a massive binary is required in the
``transitional case'' we  suggested above to explain \src\
properties.

Comparison of the 2011 spectral results with the previous \xmm\
observation shows a lower absorbing column density in 2011 than
in 2004. However, in both occasions, the absorption was well in
excess with respect to the total Galactic column density
towards the source. It is possible that the variability of the
absorption in the two observations is due to the different
orbital phases covered by the two observations, but the
uncertainties on the orbital parameters do not allow us to
determine the absolute orbital phase covered by two spectra.
The local absorbing column density due to the supergiant wind
in a HMXBs is expected to vary along the orbit, even in case of
a spherically symmetric wind, also in dependence of the
inclination of the system. Further variability can be due to
the presence of different gas structures (accretion wake 
around the NS together with tidal effects which
induce a gas stream from the supergiant) which form because of
the presence of the NS gravitational field \citep{BSK1991}.

Another important result is that the possible hint for a soft
excess present in 2004, is now well established by the 2011
observation and it is compatible with the presence of a ionized
absorber.
In particular, the cumulative X--ray spectrum during the flares
observed in 2011 (F) is well described introducing an
additional ionized absorber, resulting in an absorber
ionization state, $\xi$,  of 125~$^{+60}
_{-45}$~erg~cm~s$^{-1}$, and an absorbing column density,
N$_{\rm H absori}$, of $15.8  ^{+2.0}
_{-1.8}$$\times$$10^{22}$~cm$^{-2}$.
In the likely hypothesis of the wind photo-ionization, the
above value of the ionization parameter $\xi$ indicates, for
example, the presence of highly ionized oxygen  (O~{\sc viii})
and neon (Ne~{\sc x} ions) \citep{Kallman1982}, whose lines
could have been   observed in the RGS energy range, if the
source had been brighter and less absorbed.
The average X--ray flux during flares, corrected for both
interstellar and local (ionized) absorption is
5$\times10^{-11}$~erg~cm$^{-2}$~s$^{-1}$ (1--10 keV), which
implies a ionizing X--ray luminosity L=$10^{36}$~erg~s$^{-1}$.
Although the  $\xi$ value we obtained  is very likely an
average value, it is possible to calculate the distance, R,
from the X--ray source of the main component of the absorbing
ionized material.
Using the measured N$_{\rm H absori}$ value and assuming
N$_{\rm H absori}$ = nR, we obtain a distance
R=5$\times$$10^{10}$~cm, which is  compatible with the NS
accretion radius, and well within the orbital separation
($\sim$10$^{12}$~cm) of the binary system.

	\section{Conclusion \label{concl}}

The new \xmm\ observations we have reported here
allowed us to perform an in-depth investigation of the transient source \src.
A high dynamic range has been observed for the first time, of about
two orders of magnitude, leading to a firmer classification of \src\
as a member of the class of the SFXTs.

We obtained a more precise value of the  pulse period and we
clearly established the presence of a soft X--ray excess in the
flares X--ray spectrum, which we interpreted as due to the
presence of ionized wind material, mainly located at a distance
compatible with the NS accretion radius. The absorbing column
density variations between 2004 and 2011 are likely due to the
different orbital phases covered between the two observations.
The low intensity emission is both softer and more absorbed
than during bright flares.

The kind of X--ray variability (its high dynamic range on the observed time scale),
together with the candidate optical counterpart, the short orbital period
and the quasi-periodic flaring activity, all suggest that the X--ray emission from \src\
is driven by a transitional accretion regime, intermediate between ``pure'' wind accretion
and ``full'' Roche lobe overflow.

In conclusion, we suggest here for the first time that this hypothesis could
explain both \src\ X--ray behaviour and possibly other SFXTs with similarly short orbital periods.

\section*{Acknowledgments}

This work is based on data from observations with \xmm.
\xmm\ is an ESA science mission with instruments and
contributions directly funded by ESA Member States and the USA (NASA).
We thank the \xmm\ duty scientists and science planners
for making these observations possible,
in particular Rosario Gonzalez-Riestra (\xmm\ Science Operations Centre User Support Group).
Lara Sidoli thanks John Blondin for interesting discussions.
We made use of HEASARC online services, supported by NASA/GSFC.
%
This work was supported by the grant from PRIN-INAF 2009, ``The
transient X--ray sky: new classes of X--ray binaries containing neutron stars''
(PI: L. Sidoli).

\bibliographystyle{mn2e}

\bsp

\label{lastpage}

\end{document}